# Identity Elements of Archaeal tRNA

Bibekanand Mallick,[1] Jayprokas Chakrabarti,[1,2,*] Satyabrata Sahoo,[1] Zhumur Ghosh,[1] and Smarajit Das[1]

*Computational Biology Group (CBG), Theory Department, Indian Association for the Cultivation of Science, Calcutta 700032 India*[1] *and Biogyan, BF 286, Salt Lake, Calcutta 700064 India*[2]

**Abstract**

Features unique to a transfer-RNA are recognized by the corresponding tRNA-synthetase. Keeping this in view we isolate the discriminating features of all archaeal tRNA. These are our identity elements. Further, we investigate tRNA-characteristics that delineate the different orders of Archaea.

## 1. Introduction

Transfer RNA (tRNA) genes form the single largest gene family. The yeast $tRNA^{Ala}$ was the first to be sequenced.[1] tRNA consists of ~76 nucleotides. They are found in cytoplasm, as well as in mitochondria and chloroplast.[2,3] We deal with cytoplasmic tRNA genes of Archaea.[4] As of now, several powerful tRNA gene identification routines[5–11] are accessible. tRNA sequence databases[12,13] have grown. Archaeal tRNA genes have introns. Some introns occur at what are called canonical (i.e. between 37 and 38) sites. Introns appear at non-canonical sites (sites other than canonical) as well. The introns at non-canonical sites vary in size and number. These non-canonical introns make the detection of tRNA genes tricky. The present tRNA gene identification routines sometimes cannot sort out these introns. Consequently, some tRNA genes are embedded (http://arxiv.org/abs/q-bio/0507008), some missed or misidentified, e.g. *Nanoarchaeum equitans*. The present routines[8,11] do not get $tRNA^{Trp}$(CCA). Instead it gets a second copy of $tRNA^{Ser}$(CGA) lying between nucleotides 151 992 and 152 081. We find this to be a misidentification (http://arxiv.org/abs/q-bio.GN/0504034). We infer that the nucleotides between 151 992 and 152 078 constitute $tRNA^{Trp}$(CCA) gene. It has a non-canonical intron, 13 bases long, between positions 30 and 31 of $tRNA^{Trp}$(CCA) gene. Again, the present routines miss $tRNA^{iMet}$(CAT) gene in *N. equitans*; we find that it is located between nucleotides 35 249 and 35 429 with two non-canonical introns. For archaeal tRNA genes there are a few such cases. In this paper we address the identity elements (for recognition of cognate amino acids) in cytoplasmic tRNAs of Archaea.

In this study, we have analyzed 22 species of 12 orders of three phyla of Archaea [the sequences deposited at DDBJ (www.ddbj.nig.ac.jp/), NCBI, DOGAN (http://www.bio.nite.go.jp/dogan/Top/) and Microbial Genome Database (MGBD) at http://mbgd.genome.ad.jp/] including the recent phylum Nanoarchaeota and recently sequenced *Picrophilus torridus.* Out of these 22 species, 3 species whose sequencing are not yet completed, but the draft genome sequence is deposited by DOE Joint Genome Institute (JGI) (http://genome.ornl.gov/microbial/) are also taken for analysis. These three draft sequence are of *M. barkeri* and *M. burtonii* (Methanosarcinales) and *F. acidarmanus* (Thermoplasmatales).

Table 1 lists the Archaea and the abbreviations used to denote them. The other notations used in this paper are as follows: (i) $tRNA^{Trp}$(CCA) means tryptophan tRNA with anticodon, CCA; (ii) G1:C72 means base pairing between first nucleotide and 72nd nucleotide of tRNA.

* To whom correspondence should be addressed. Tel. +91 33 24734971, ext. 281 (Off.). Fax. +91 33 24732805; E-mail: tpjc@iacs.res.in or biogyan@vsnl.net

**Table 1.** Various features of 22 species of Archaea.

| Name of Archaeal species | Abbreviated names | No. of tRNAs | Optimal growth temp. (°C) | Features |
|---|---|---|---|---|
| **Crenarchaeota** | | | | |
| *Pyrobaculum aerophilum*[32] (NC_003364) | Paero | 46 | 100 | Facultative nitrate reducing anaerobe and hyperthermophile |
| *Aeropyrum pernix*[33] (NC_000854) | Aper | 47 | 90 | Aerobic chemorganotroph; sulfur enhances growth and hyperthermophile |
| *Sulfolobus solfataricus*[34] (NC_002754) | Ssolfa | 46 | 80 | Aerobe, metabolizes sulfur; thermoacidophile |
| *Sulfolobus tokodaii*[35] (NC_003106) | Stok | 46 | 80 | Aerobe, metabolizes sulfur; thermoacidophile |
| **Euryarchaeota** | | | | |
| *Archaeoglobus fulgidus*[36] (NC_000917) | Afulg | 46 | 83 | Anaerobic, sulfate reducing, motile |
| *Halobacterium* sp. NRC-1[37] (NC_002607) | Halo | 47 | 37 | Aerobic, obligate halophiles, with a cell envelope, two extra chromosomal element, motile |
| *Pyrococcus furiosus* DSM 3638[38] (NC_003413) | Pf | 46 | 96 | Anaerobic heterotroph, sulfur enhances growth, motile |
| *Pyrococcus horikoshii*[39] (NC_000961) | Phori | 46 | 96 | Anaerobic heterotroph, sulfur enhances growth, motile |
| *Pyrococcus abyssi*[40] (NC_000868) | Paby | 46 | 96 | Anaerobic heterotroph, sulfur enhances growth, motile |
| *Methanococcus maripaludis* S2[41] (NC_005791) | Mmari | 37 | 85 | Strictly anaerobe, fix nitrogen |
| *Methanocaldococcus jannaschii*[42] (NC_000909) | Mjan | 36 | 85 | Strict anaerobe, methanogen, motile; two extra chromosomal elements |
| *Methanosarcina mazei* Goe1[43] (NC_003901) | MmGoe | 57 | 37 | Anaerobe, possibly capable of aerobic growth; nitrogen fixing, versatile methanogen; motile |
| *Methanosarcina acetivorans*[44] (NC_003552) | Macet | 55 | 37 | Anaerobe, possibly capable of aerobic growth; nitrogen fixing, versatile methanogen; motile |
| *Methanosarcina barkeri*[a] | Mbar | 59 | 37 | Anaerobe, versatile methanogen, found in oil wells, sewage and rumen of ungulates |
| *Methanococcoides burtonii*[a] | Mburt | 44 | 23 | Psychrotolerant methanogen |
| *Thermoplasma acidophilum*[45](NC_002578) | Tacid | 46 | 59 | Facultative anaerobe, thermoacidophilic, aerobically able to metabolize sulfur, motile with PM |
| *Thermoplasma volcanium*[46] (NC_002689) | Tvol | 46 | 60 | Facultative anaerobe, thermoacidophilic, aerobically able to metabolize sulfur, motile with PM |
| *Picrophilus torridus* DSM 9790[47] (NC_005877) | Pitor | 46 | 60 | Aerobic, thermoacidophile, found in dry solfataric land, cell wall present |
| *Ferroplasma acidarmanus*[a] | Facid | 44 | 35 | Aerobic, acidophilic, heterotroph |
| *M.thermautotrophicus*[48] (NC_000916) | Mther | 39 | 65 | Strict anaerobe, nitrogen fixing methanogen |
| *Methanopyrus kandleri* AV19[49] (NC_003551) | Mkan | 35 | 110 | strict anaerobe, methanogen, with high cellular salt concentration |
| **Nanoarchaeota** | | | | |
| *Nanoarchaeum equitans* Kin4-M[50] (NC_005213) | NanoA | 41 | 70–98 | Hyperthermophile, obligate symbiotic |

PM means plasma membrane.
[a] denotes the draft genomes whose whole genome sequencing is going on.

The abbreviation bp stands for base pair/base pairs. Semi-conserved bases are the ones that are conserved with but a few exceptions. 3D bp means three-dimensional base pairings. These are essential for tertiary structure.

The primary tRNA nucleotide chain is single stranded. This chain folds back onto itself to form the base paired secondary structure. This folded cloverleaf structure confers on tRNA some of its functions. The secondary structure of tRNA[14] has (i) Acceptor arm or A-arm: In this, the 5′ and 3′ ends of tRNA are base paired into a stem of 7 bp. (ii) DHU arm or D-Arm: Structurally a stem-loop, D-arm frequently contains the modified base dihydrouracil. (iii) Anticodon arm or AC-arm, made of a stem and a loop containing the anticodon. The canonical structure of AC-loop is essential for interactions with ribosomal A and P sites.[15] At 5′ end of this loop is a pyrimidine base at 32, followed by an invariant U at 33. The anticodon triplet, at 34, 35, 36 is in the exposed loop region.[13] (iv) Extra arm, or V-arm: This arm is not always present. It is of variable length and is largely responsible for variations in the lengths of tRNAs. The classification of tRNAs into types I and II depends on the length of V-arm. (v) T-$\psi$-C Arm or T-arm: This arm has conserved sequence of three ribonucleotides: ribothymidine, pseudouridine and cytosine. T-arm has stem–loop secondary structure and (vi) tRNA terminates with CCA at the 3′ end.[16] For tRNA genes, CCA may not be present.

In case CCA is absent in tRNA genes, it is added during the maturation to tRNA.

The attachment of cognate amino acids to their corresponding tRNA are catalyzed by aminoacyl-tRNA synthetase (AARS).[17] Accurate acylation of tRNA depends on two factors: a set of nucleotides in tRNA molecule (identity elements) responsible for proper identification by AARS[18–21] and competition between different synthetases for tRNAs.[22–25] Tertiary L-shape of tRNA facilitates its identification by AARS for aminoacylation.[26] L-shape comes about through the interactions between D-arm and T-arm. There are a few key features that maintain the L-shape of tRNA.[27,28] These interactions include Watson–Crick base pairing, Hoogsteen base pairing, and triple-helical base pairing. It is generally accepted that the major interactions maintaining the L-shape occur at the corner of the molecule where D- and T-loops meet. This region, called DT, contains several elements, including the reverse-Hoogsteen base pair U54:A58 and C55-mediated U-turn in T-loop, the inter-loop base pairs G18:C55 and G19:C56 and stack of four mutually intercalated purine bases A58–G18–R57–G19.[29,30] This intraloop U54:A58 is stacked on G53.C61 at the end of T stem and forces the two bases at positions 59 and 60 to loop out, forming a characteristic T-loop of five bases instead of seven. This characteristic T loop conformation is important for recognition by elongation factors.[31] We propose here all possible identity elements in tRNAs of Archaea for recognition of cognate amino acids. These are bases within tRNA sequences with which it would be possible to identify each of the tRNAs of individual archaeal species of different taxonomic groups.

From the 22 archaeal genomes, we sourced **1001** tRNA genes in total. We located tRNA genes using our in-house algorithm. Subsequently, we ran tRNAScan-SE and ARAGORN as additional checks.

## 2. Materials and Methods

The archaeal genomes are obtained from NCBI and Genome Information Broker (http://gib.genes.nig.ac.jp/). Raw tRNA gene sequences are found by searching the different motifs present in the consensus sequence of different tRNAs genes of Archaea. At first we adopted the standard cloverleaf model[1] for studying the secondary structure of predicted tRNA genes of Archaea. In doing so, we got some false positives and a few tRNA genes were missed out. We then imposed constraints, unique to archaeal tRNA genes. A regular cloverleaf structure was searched in tRNA genes of the genomes by adopting archaeal tRNA gene features. The constraints of lengths of stems of regular tRNA A-arm, D-arm, AC-arm and T-arm are 7, 4, 5 and 5 bp, respectively. In a few cases, the length of D-arm and AC-arm are relaxed. In addition, parameters and constraints used in the search for cloverleaf tRNAs are the following: (i) T8 (except Y8 in *M. kandleri*), G18, R19, R53, Y55 and A58 are considered as conserved bases for Archaea. (ii) The lengths of introns and V-arm are allowed from 6 to 121 and up to 21, respectively. (iii) Positions optionally occupied in D-loop are 17, 17a, 20a and 20b. (iv) Canonical and non-canonical introns may or may not be present. Keeping these constraints, we were able to extract tRNA genes in all Archaea. (Number of tRNA genes obtained are given in Table 1). After getting the tRNA genes we ran the standard routines[8,11] to check for the secondary structure.

## 3. Results and Discussion

Table 2 lists the conserved and semi-conserved bases/base pairs of various archaeal elongator tRNAs. In Table 3 we list the following identity elements.

(i) *tRNA*$^{Ala}$. G3:U70 is a unique base pair in archaeal tRNA$^{\mathrm{Ala}}$. Ala-RS specifically aminoacylates tRNA$^{\mathrm{Ala}}$ because of direct recognition of unique functional groups exposed on G3:U70. The G3:U70 pair contributes to tRNA$^{\mathrm{Ala}}$ identity by three mechanisms. First, the wobble pair presents a distinctive array of hydrogen bond donors and acceptors in major and minor grooves. Second, unusual structural features induced by the G3:U70 pair may influence Ala-RS binding to AC-stem and/or 3′-CCA end of tRNA$^{\mathrm{Ala}}$. Third, G3:U70 may destabilize AC-end and favor the formation of an optimal active site geometry by induced fit.[51] These mechanisms are not mutually exclusive. The role of the functional groups in minor groove of A-stem on recognition and aminoacylation by Ala-RS has been extensively investigated.[52] A73 discriminator base could be another identity element.

(ii) *tRNA*$^{Arg}$. 1:72 of tRNA$^{\mathrm{Arg}}$ is conserved for Archaea, except A:U for tRNA$^{\mathrm{Arg}}$(CCG) of MmGoe, Macet and Mbar. The third base of anticodon G36 or U36 and the second anticodon base C35 are arginine identities. A20 and the discriminator base A73 or G73 are identity elements as well. Nanoarchaea has U20 and A59. These unusual nucleotide positions may be identity elements for tRNA$^{\mathrm{Arg}}$ of Nanoarchaea.

(iii) *tRNA*$^{Asn}$. Methanosarcinales have unique A10:U25 unlike G10:C25 in other Archaea. Similarly, all three Thermococcales (Pf, Paby and Phori) and Afulg, all hyperthermophiles, have U13:G22 in tRNA$^{\mathrm{Asn}}$. Other Archaea have C13:G22. For archaeal tRNA$^{\mathrm{Asn}}$, we find that C3:G70, G73 and G34, U35, U36 are the identity elements.

(iv) *tRNA*$^{Asp}$. G4:C69 is characteristic of Crenarchaea which distinguishes it from Euryarchaea that have C4:G69 (exceptions: C4:G69 for Paero and U4:A69 in Pitor). Methanosarcinales differ at U2:G71 from

**Table 2.** Conserved and semi-conserved bases/base pairs of various Archaeal elongator tRNAs.

| Position | Ala CO | Ala EO | Ala NO | Arg CO | Arg EO | Arg NO | Asn CO | Asn EO | Asn NO | Asp CO | Asp EO | Asp NO | Cys CO | Cys EO | Cys NO | Gln CO | Gln EO | Gln NO | Glu CO | Glu EO | Glu NO | Gly CO | Gly EO | Gly NO | His CO | His EO | His NO | Ile CO | Ile EO | Ile NO |
|---|---|---|---|---|---|---|---|---|---|---|---|---|---|---|---|---|---|---|---|---|---|---|---|---|---|---|---|---|---|---|
| **A-arm** | | | | | | | | | | | | | | | | | | | | | | | | | | | | | | |
| 1–72 | GC | GC | GC | GC | | GC | GC | GC | GC | GC | GC | GC | GC | GC | GC | AU | AU | AU | GC | GC | GC | GC | | GC | GC | GC | | GC | GC | GC |
| 2–71 | GC | GC | GC | GC | | GC | CG | CG | CG | CG | | CG | CG | | CG | GC | GC | GC | CG | CG | CG | CG | CG | CG | CG | CG | | GC | GC | GC |
| 3–70 | GU | GU | GU | | | GC | CG | CG | CG | CG | CG | CG | CG | | CG | CG | | CG | CG | | CG | GC | GC | GC | | CG | | GC | GC | GC |
| 4–69 | CG | CG | CG | CG | | CG | GC | GC | CG | GC | CG | CG | GC | | GC | CG | CG | CG | GC | CG | GC | GC | | CG | CG | GC | | CG | | CG |
| 5–68 | CG | | CG | CG | | CG | | CG | GC | CG | | CG | GC | | CG | GC | CG | CG | CG | CG | CG | CG | | CG | GC | GC | | CG | | CG |
| 6–67 | GC | | CG | CG | | | | | GC | GC | GC | GC | GC | GC | CG | GC | | CG | GC | GC | CG | CG | | CG | GC | GC | | CG | | CG |
| 7–66 | GC | | GC | GC | | | GC | | GC | GC | | GC | GC | | GC | GC | GC | GC | GC | | GC | GC | | GC | GC | | | GC | GC | GC |
| **3D-bp** | | | | | | | | | | | | | | | | | | | | | | | | | | | | | | |
| 8–14 | UA | UA | UA | UA | UA | UA | UA | UA | UA | UA | UA | UA | UA | UA | UA | UA | UA | UA | UA | UA | UA | UA | UA | UA | UA | UA | | UA | UA | UA |
| 15–48 | GC | GC | GC | GC | | GC | GC | GC | GC | GC | GC | GC | GC | GC | GC | GC | GC | GC | GC | GC | GC | GC | GC | GC | GC | GC | | GC | GC | GC |
| 18–55 | GU | GU | GU | GU | GU | GU | GU | GU | GU | GU | GU | GU | GU | GU | GU | GU | GU | GU | GU | GU | GU | GU | | GU | GU | GU | | GU | GU | GU |
| 19–56 | GC | GC | GC | GC | GC | GC | GC | GC | GC | GC | GC | GC | GC | GC | GC | GC | GC | GC | GC | GC | GC | GC | GC | GC | GC | GC | | GC | GC | GC |
| 26–44 | GA | GA | GA | | | GA | GU | | GA | GU | | GC | GU | | GU | | | GU | GU | | GA | | | GU | GA | | | GA | | GA |
| 32–38 | | | CA | CA | CA | CA | CA | CA | CA | CA | | CA | CA | CA | CA | | CA | CU | CA | | AA | | | CA | CA | CA | | CA | | CA |
| 54–58 | UA | UA | UA | UA | UA | UA | UA | UA | UA | UA | UA | UA | UA | UA | UA | UA | UA | UA | UA | UA | UA | UA | UA | UA | UA | UA | | UA | UA | UA |
| **D-stem** | | | | | | | | | | | | | | | | | | | | | | | | | | | | | | |
| 10–25 | GU | GC | GC | GC | | GC | GC | | GC | GU | GU | GU | GC | GC | CG | GU | GU | GU | GU | GU | GC | GC | GC | GU | GC | | | GC | GC | GC |
| 11–24 | CG | | UA | CG | | CG | CG | CG | CG | UA | UA | UA | CG | CG | CG | UA | | UA | UA | UA | CG | UA | UA | UA | CG | | | CG | CG | CG |
| 12–23 | UA | UA | CG | UA | | UA | UA | UA | UA | | GC | GC | | | CG | CG | GC | GC | | GC | UA | CG | | CG | UA | GC | | UA | UA | UA |
| 13–22 | CG | CG | CG | CG | | | CG | | CG | UG | ---- | UG | | | ---- | UG | | UG | UG | | CG | UG | | UG | CG | | | | | CG |
| **AC-stem** | | | | | | | | | | | | | | | | | | | | | | | | | | | | | | |
| 27–43 | CG | | CG | CG | | | CG | | GC | | | CG | GU | | UA | CG | | CG | CG | | CG | CG | | CG | | | | CG | CG | CG |
| 28–42 | | | CG | CG | | CG | CG | | CG | GC | | GC | CG | | CG | | | GC | GC | | CG | CG | | GC | CG | | | CG | | CG |
| 29–41 | CG | | GC | GC | | GC | | | CG | GC | GC | GC | GC | | GC | GC | | GC | GC | GC | CG | GC | | GC | GC | GC | | CG | CG | CG |
| 30–40 | GC | | CG | GC | | GC | GC | GC | GC | GC | | GC | GC | GC | GC | GC | GC | GC | GC | GC | GC | GC | | GC | GC | | | GC | GC | GC |
| 31–39 | | | CG | CG | | | GC | | GC | CG | | CG | GC | | GC | GC | | GC | CG | | GC | CG | | CG | | | | GC | | GC |
| **T-stem** | | | | | | | | | | | | | | | | | | | | | | | | | | | | | | |
| 49–65 | CG | | CG | | | GC | | | CG | CG | | CG | | | GC | CG | | CG | CG | | GC | CG | | CG | CG | CG | | | | |
| 50–64 | CG | | CG | CG | | CG | | | CG | CG | CG | CG | CG | | GC | | | CG | CG | | CG | CG | CG | CG | CG | CG | | | | |
| 51–63 | GC | GC | GC | GC | GC | GC | | | CG | GC | GC | GC | GC | | GC | GC | | CG | GC | GC | GC | GC | GC | GC | GC | GC | | GC | | GC |
| 52–62 | GC | GC | GC | GC | | GC | GC | GC | GC | GC | GC | GC | GC | GC | GC | GC | | GC | GC | GC | GC | GC | GC | GC | GC | | | GC | GC | GC |
| 53–61 | GC | GC | GC | GC | | GC | GC | GC | GC | GC | GC | GC | GC | GC | GC | GC | GC | GC | GC | GC | GC | GC | GC | GC | GC | GC | | GC | GC | GC |
| 33 | U | U | U | U | U | U | U | U | U | U | U | U | U | U | U | U | U | U | U | U | U | U | U | U | U | U | | U | U | U |
| 73 | A | A | A | G | | G | G | G | G | G | G | G | U | U | U | A | A | A | A | A | A | A | A | A | C | C | | A | A | A |
| 9 | A | A | A | A | -- | A | A | A | A | A | -- | G | | | G | C | -- | G | | --- | A | --- | --- | A | A | -- | | A | -- | A |

| Position | Amino acids | | | | | | | | | | | | | | | | | | | | | | | | | | | | | |
|---|---|---|---|---|---|---|---|---|---|---|---|---|---|---|---|---|---|---|---|---|---|---|---|---|---|---|---|---|---|---|
| | Leu | | | Lys | | | Phe | | | Pro | | | Ser | | | Thr | | | Trp | | | Tyr | | | Val | | | eMat | | |
| | CO | EO | NO | CO | EO | NO | CO | EO | NO | CO | EO | NO | CO | EO | NO | CO | EO | NO | CO | EO | NO | CO | EO | NO | CO | EO | NO | CO | EO | NO |
| **A-arm** | | | | | | | | | | | | | | | | | | | | | | | | | | | | | | |
| 1–72 | GC | GC | GC | GC | GC | GC | GC | GC | GC | GC | GC | GC | GC | GC | GC | GC | GC | GC | GC | GC | GC | CG | CG | CG | GC | GC | GC | GC | GC | GC |
| 2–71 | CG | CG | CG | GC | GC | GC | CG | CG | GC | GC | GC | GC | CG | | CG | CG | CG | CG | GC | GC | GC | CG | CG | CG | GC | GC | GC | ** | ** | CG |
| 3–70 | GC | | GC | GC | GC | GC | CG | CG | GC | GC | GC | GC | CG | | CG | CG | CG | CG | GC | GC | GC | CG | CG | GC | GC | GC | GC | ** | ** | CG |
| 4–69 | GC | | GC | CG | | CG | GC | | GC | GC | | CG | GC | GC | GC | GC | | GC | CG | GC | GC | GC | GC | GC | CG | CG | CG | ** | | GC |
| 5–68 | GC | | CG | CG | CG | CG | CG | | CG | CG | | CG | GC | | CG | CG | | CG | CG | | CG | CG | | GC | CG | | CG | CG | | CG |
| 6–67 | GC | | CG | CG | | CG | CG | GC | GC | CG | | CG | GC | GC | CG | | | CG | CG | CG | CG | CG | | CG | CG | | CG | CG | | CG |
| 7–66 | GC | GC | GC | GC | GC | GC | GC | | GC | GC | GC | GC | GC | | GC | GC | GC | GC | GC | | CG | GC | | GC | GC | GC | GC | GC | | GC |
| **3D-bp** | | | | | | | | | | | | | | | | | | | | | | | | | | | | | | |
| 8–14 | UA | UA | UA | UA | UA | UA | UA | UA | UA | UA | UA | UA | UA | UA | UA | UA | UA | UA | UA | UA | UA | UA | UA | UA | UA | UA | UA | UA | UA | UA |
| 15–48 | GC | GC | GC | GC | GC | GC | GC | GC | GC | GC | GC | GC | GC | GC | GC | GC | GC | GC | GC | | GC | GC | GC | GC | GC | GC | GC | GC | GC | GC |
| 18–55 | GU | GU | GU | GU | GU | GU | GU | GU | GU | GU | GU | GU | GU | GU | GU | GU | GU | GU | GU | GU | GU | GU | GU | GU | GU | GU | GU | GU | GU | GU |
| 19–56 | GC | GC | GC | GC | GC | GC | GC | GC | GC | GC | GC | GC | GC | GC | GC | GC | GC | GC | GC | GC | GC | GC | GC | GC | GC | GC | GC | GC | GC | GC |
| 26–44 | GU | GU | GU | GA | | GA | GU | | GA | GU | | GA | GU | GU | GU | | | GA | | | GA | GU | | GA | GA | | GA | GU | | GA |
| 32–38 | | | CA | CA | CA | CA | CA | CA | CA | | | UU | CA | | CA | CA | | CA | CA | CA | CA | CA | CA | CA | CC | | CA | CA | CA | CA |
| 54–58 | UA | UA | UA | UA | UA | UA | UA | UA | UA | UA | UA | UA | UA | UA | UA | UA | UA | UA | UA | UA | UA | UA | UA | UA | UA | UA | UA | UA | UA | UA |
| **D-stem** | | | | | | | | | | | | | | | | | | | | | | | | | | | | | | |
| 10–25 | CG | | CG | GC | GC | GC | GC | GC | GC | GU | GU | GC | | GC | CG | GC | GC | GC | | GU | GC | GC | GC | GC | GC | | GC | GC | GC | GC |
| 11–24 | CG | CG | CG | CG | CG | CG | CG | CG | CG | UA | GC | UA | CG | CG | CG | CG | CG | CG | CG | | CG | CG | CG | CG | UA | UA | UA | CG | CG | CG |
| 12–23 | CG | | CG | UA | UA | UA | UA | UA | UA | CG | GC | CG | CG | | CG | UA | UA | UA | UA | GC | UA | UA | UA | UA | CG | CG | CG | UA | | UA |
| 13–22 | ----- | ----- | ----- | CG | | CG | CG | CG | CG | UG | ---- | | ---- | | --- | CG | | CG | CG | --- | CG | CG | CG | CG | UG | ---- | ---- | | | CG |
| **AC stem** | | | | | | | | | | | | | | | | | | | | | | | | | | | | | | |
| 27–43 | | | UA | GU | | CG | CG | CG | CG | CG | | CG | | | | CG | | GC | | | GU | CG | | GC | CG | | CG | CG | CG | CG |
| 28–42 | | | CG | CG | | CG | CG | | CG | | | GC | | | | CG | | CG | CG | | CG | CG | | CG | CG | CG | CG | CG | | CG |
| 29–41 | GC | | GC | GC | | GC | CG | | CG | | | GC | GC | | | GC | | GC | GC | GC | GC | CG | | CG | GC | | GC | | | CG |
| 30–40 | GC | GC | GC | GC | GC | GC | GC | | GC | GC | | GC | GC | GC | GC | | | GC | GC | GC | GC | GC | GC | GC | CG | CG | CG | GC | GC | GC |
| 31–39 | | | | GC | | GC | GC | | GC | | | | | | GC | CG | | CG | GC | | ---- | GC | AU | GC | CG | CG | CG | GC | | GC |
| **T-stem** | | | | | | | | | | | | | | | | | | | | | | | | | | | | | | |
| 49–65 | GC | | CG | CG | | GC | CG | | GC | CG | | CG | GC | | GC | CG | | GC | GC | | GC | CG | | GC | CG | | CG | CG | | GC |
| 50–64 | | | GC | CG | | CG | GC | | GC | CG | | CG | CG | | GC | CG | | CG | | | GC | GC | CG | GC | | | GC | | | CG |
| 51–63 | GC | GC | GC | GC | GC | GC | GC | | GC | GC | | GC | GC | GC | GC | | GC | GC | GC | GC | GC | GC | | GC | GC | | GC | GC | | GC |
| 52–62 | GC | | GC | GC | GC | GC | GC | GC | GC | GC | | GC | GC | | GC | GC | | GC | GC | GC | GC | GC | | GC | GC | | GC | GC | | GC |
| 53–61 | GC | GC | GC | GC | GC | GC | GC | GC | GC | GC | GC | GC | GC | GC | GC | GC | GC | GC | GC | GC | GC | GC | GC | GC | GC | GC | GC | GC | GC | GC |
| 33 | U | U | U | U | U | U | U | U | U | U | U | U | U | U | U | U | U | U | U | U | U | U | U | U | U | U | U | U | U | U |
| 73 | A | A | A | G | G | G | A | A | A | A | A | A | G | G | G | U | U | U | A | A | A | A | A | A | A | A | A | A | A | A |
| 9 | G | -- | G | A | A | A | A | A | A | --- | --- | A | G | --- | G | --- | --- | A | A | G | A | A | A | A | C | --- | A | A | A | A |

Conserved and semi-conserved bases/bp are marked in black and red font respectively.
Dashed lines (colour indication is same as above) indicate no base pairing.
**denotes interchange of GC and CG bp at a particular bp position.

**Table 3.** Identity elements of Archaeal tRNAs.

| tRNAs | Identity elements | Comments |
|---|---|---|
| Ala | (a) A73 | |
| | (b) G3:U70 | |
| Arg | (a) A/G73 | For NanoA,U20 and A59 are |
| | (b) G36 or U36 and C35 | Identity elements |
| | (c) A20 | |
| Asn | (a) G73 | |
| | (b) G34, U35, U36 | |
| | (c) C3:G70 | |
| Asp | (a) G73 | |
| | (b) G34,U35,C36 | |
| | (c) G6:C67 | |
| Cys | (a) U73 | |
| | (b) G34,C35,A36 | |
| Gln | (a) A73 | For eMet, Identity elements are |
| | (b) Anticodon and specially modified U at 34th position | A73 & C34, A35 and U36 |
| | (c) A1:U72 | |
| Glu | (a) U34,U35,C36 | |
| | (b) C5:G68 | |
| Gly | (a) A73 | |
| | (b) C35,C36 | |
| His | (a) C73 | |
| | (b) C50:G64, G29:C41 | |
| Ile | (a) A73 | |
| | (b) G34,A35,U36 | |
| | (c) C27:G43 | |
| Leu | (a) A73 | |
| | (b) G26:U44 | |
| Lys | (a) G73 | |
| | (b) U35 | |
| iMet | (a) A1:U72 | |
| | (b) Three consecutive GC at bottom of AC-stem | |
| Phe | (a) A73 | |
| | (b) G34, A35,A36 | |
| | (c) C13:G22 | |
| Pro | (a) A73 | |
| | (b) G36 | |
| | (c) Three GC at 5′ end of A-arm | |
| Ser | (a) G73 | |
| | (b) G26:U44 | Additionally, V-arm plays a |
| Thr | (a) U73 | role as Identity element |
| | (b) G35, U36 | |
| Trp | (a) A73 | |
| | (b) C34,C35,A36 | |
| Tyr | (a) C1:G72 | |
| | (b) G34,U35 and A36 | |
| Val | (a) A73 | |
| | (b) A35, C36 | |

Met is for initiator tRNA and eMet is for elongator methionine tRNA.

C2:G71 in other Archaea. We surmise G6:C67 to be the identity determinant recognized by synthetase in addition to other identity elements of $tRNA^{Asp}$ in its anticodon, G34, U35 and C36.

(v) $tRNA^{Cys}$. Thermoplasmales differs from other archaeal order at G2:C71 and A3:U70. Others have CG at these two positions. $tRNA^{Cys}$(GCA) of Nanoarchaea differs from two other archaeal phyla at C10:G25. The two other phyla have G10:C25.

(vi) $tRNA^{Gln}$. C12:G23 in D-stem of Crenarchaea is a distinguishing feature of $tRNA^{Gln}$ from G12:C23 of Euryarchaea. A at 73rd position is the discriminator base, with exception of $tRNA^{Gln}$(UUG) of Facid where G is the discriminator. The other identity elements of $tRNA^{Gln}$ could be A1:U72 in addition to anticodon bases at 34th, 35th and 36th positions. The modified U base at 34th nucleotide could be the major identity element for the synthetase as in *E.coli*.[53]

(vii) $tRNA^{Glu}$. The 3D bp 15:48 is GC, but $tRNA^{Glu}$(UUC) of Mkan has CC at this position. The Sulfolobales and Desulfurococcales $tRNA^{Glu}$ differ from all Crenarchaea and Euryarchaea with G4:C69 from C4:G69 in all others. U34, U35 and C36 are identity elements for Archaea as in *Escherichia coli*.[54,55] C5:G68 could be another identity element for archaeal $tRNA^{Glu}$.

(viii) $tRNA^{Gly}$. Nanoarchaea is unique at 33rd site in its $tRNA^{Gly}$. It has pyrimidine C in $tRNA^{Gly}$(CCC) rather than U as in other two phyla. A73 acts as an identity element in addition to C35, C36.

(ix) $tRNA^{His}$. Methanococcales, Methanosarcinales and Thermoplasmales have G7:U66 not observed in other Archaea. Paero has UA at 11:24 unlike other Crenarchaea where it is CG. The discriminator base C73, the fourth unpaired base from the tRNA 3′ end are the characteristic features of $tRNA^{His}$(GUG). These are the identity elements in addition to the anticodon. C50:G64 could be an additional identity element. G29:C41 could be a minor identity element as well.

(x) $tRNA^{Ile}$. Nanoarchaea unusually have $tRNA^{Ile}$(UAU) and $tRNA^{Ile}$ (GAU) which differ from each other at 29:41, 31:39 and 49:65 with interchange of GC or CG. The first one (i.e. $tRNA^{Ile}$(UAU)) located between bases 225 624 and 225 716 has an intron unlike the second and it is the first $tRNA^{Ile}$(UAU) to be identified in Archaea. The two $tRNA^{Ile}$(GAU) of Macet are identical and have 3′-CCA arm. C27:G43 is presumably the identity element for $tRNA^{Ile}$(GAU). In addition, the anticodon and A73 are identity elements.

(xi) $tRNA^{Leu}$. MmGoe, Macet and Mburt have six $tRNA^{Leu}$ with two identical copies of $tRNA^{Leu}$(GAG). The two copies of $tRNA^{Leu}$(GAG) gene of MmGoe are identical, but differ by CCA arm in one of them. A5:U68 is characteristic of $tRNA^{Leu}$ of Thermoplasmales, except GC in $tRNA^{Leu}$(UAA). We surmise that G26:U44 is the discriminating factor (Table 3).

(xii) *tRNA$^{Lys}$*. Hyperthermophilic Desulfurococcales have C32:C38; but other Archaea have CA. The discriminator bases G73 and U35 are the acceptor identities for tRNA$^{Lys}$ as in *E coli*.[56,57]

(xiii) *tRNA$^{eMet}$ (Elongator methionine tRNA)*. Two elongator tRNA$^{Met}$(CAU) differ from one another at two positions. The first copy, tRNA$^{eMet1}$(CAU) has CG at 2:71 and 3:70 of A-arm and the second tRNA$^{eMet2}$(CAU) has GC at these positions. tRNA$^{eMet1}$(CAU) of most of the Euryarchaea have canonical introns, but rare in its second copy. C34, A35 and U36 could be the identity elements in addition to the discriminator base A73.

(xiv) *tRNA$^{iMet}$(CAU) (Initiator methionine tRNA)*. Interestingly, Mmari, MmGoe, Macet have two copies of tRNA$^{iMet}$(CAU). G28:C42 in Crenarchaeal tRNA$^{iMet}$(CAU) distinguishes it from either CG or UG at this position in Euryarchaea. Mkan is again an exception in this respect where G28:C42 is present as in Crenarchaea. tRNA$^{iMet}$ of Halo is characterized by A7:U66 in contrast to G7:C66 in other Archaea.

Archaeal tRNA$^{iMet}$(CAU) possesses unique sequence characteristics: (a) A1:U72 bp at the beginning of A-arm, (b) the three consecutive G and three C at the bottom of AC-stem forming GC bp in tRNA$^{iMet}$(CAU) of Euryarchaea similar to bacteria.[58] Four consecutive GC bps in Crenarchaeal tRNA$^{iMet}$ unlike bacteria and eukaryotes. (c) G11:C24 in D- stem in contrast to C11:G24 in elongator tRNAs, (d) UA/AU bps at 51:63 in T-stem in contrast to GC for elongator tRNAs in all Archaea. The identity elements for tRNA$^{iMet}$ are listed in Table 3.

(xv) *tRNA$^{Phe}$(GAA)*. Crenarchaeal tRNA$^{Phe}$(GAA) differs from Euryarchaea at C6:G67 against G6:C67 for Euryarchaea. Exception: halophilic tRNA$^{Phe}$(GAA) are similar at 6:67 to Crenarchaea. C13:G22, we conjecture is an identity element. In addition to this, anticodon and A73 are identity elements.

(xvi) *tRNA$^{Pro}$*. Nanoarchaea has G10:C25. It is G10:U25 in Euryarchaea and either G10:C25 or G10:U25 in Crenarchaea. tRNA$^{Pro}$ of Mther has A7:U66 in tRNA$^{Pro}$(UGG) which discriminate it from tRNA$^{Pro}$(UGG) of all other Archaea, where G7:C66 is observed. The three GC bps at 5′ end of A-arm and A73 are the most probable regions for tRNA recognition for tRNA$^{Pro}$ and G36 is a major determinant for aminoacylation, as reported for Mjan.[59] In addition to aminoacylation of tRNA$^{Pro}$ by ProRS, this AARS has a dual function of aminoacylating tRNA$^{Cys}$ in Mjan, Mmari and Mther.[60]

(xvii) *tRNA$^{Ser}$*. 6:67 is GC in Crenarchaea and Euryarchaea, but Nanoarchaea has CG. The present routines[8,11] get a second copy of tDNA$^{Ser}$(CGA) in *N. equitans* lying between neclotides 151 992 and 152 081. We find that this is a misidentification and explained in detail in tRNA$^{Trp}$(CCA) section. We find for archaeal tRNA$^{Ser}$ G26:U44 can contribute to its recognition by Ser-RS. G73 and V-loop act as identity elements as in *E. coli* tRNA$^{Ser}$.[61]

(xviii) *tRNA$^{Thr}$*. Thermococcales have conserved C12:G23 in tRNA$^{Thr}$(CGU) and tRNA$^{Thr}$(UGU) distinguishing these from other Euryarchaeal tRNA$^{Thr}$ that have UA. For tRNA$^{Thr}$(UGU),31:39 bp differentiate between Crenarchaea and Euryarchaea; these have CG (exception: Paero has A:U) and AU respectively. U73, discriminator base is involved in recognition by Thr-RS in Archaea.[62] Second and third letters of anticodon, G35 and U36 are identity elements of archaeal Thr-RS as well.

(xix) *tRNA$^{Trp}$(CCA)*. Tryptophan is a single codon amino acid and Archaea have at least one tRNA$^{Trp}$(CCA). The current standard routines[8,11] could not identify tRNA$^{Trp}$(CCA) in Nanoarchaea and we are able to identify it. The current standard routines identify the bases from 151 992 and 152 081 in Nanoarchaea as being tRNA$^{Ser}$(CGA). We argue now that the bases between 151 992 and 152 081 are unlikely to be tRNA$^{Ser}$(CGA). The reasons are as follows

(a) In tRNA$^{Ser}$ it is known that G26:U44, G73 and the elements in long V-arm are important for recognition by Ser-RS.[63] None of these recognition bases appear between 151 992 and 152 081. Instead we get G26:A44, A73 and a short V-arm.
(b) Variable arm for tRNA$^{Ser}$ is known to be long.[61] If we identify 151 992–152081 to be tRNA$^{Ser}$(CGA), it is going to have a short V-arm missing out the elements important for its recognition by Ser-RS.
(c) However, the secondary structure generated from 151 992 to 152 078 have the following important positions G1, G2, G3, A21, T33, A73, G18:T55, G19:C56, T54:A58 and G30:C40 and anticodon CCA at 34, 35 and 36. These are known to be conserved in archaeal tRNA$^{Trp}$(CCA). Therefore, the range 151 992–152 078 is tRNA$^{Trp}$(CCA) gene. It has a non-canonical intron, 13 base-long between 30 and 31. It has BHB splicing motif as well.

U12:A23 in Crenarchaea distinguishes it from either GC or CG in Euryarchaea. The ninth base discriminates between Crenarchaea and Euryarchaea; the former have A and the latter have G. The anticodons C34, C35 and A36 are recognized by Trp-RS.[64] Additionally, A73 could also be the identity element.

(xx) *tRNA$^{Tyr}$*. tRNA$^{Tyr}$(GUA) of Archaea have characteristic base pairs at 31:39:GC bias in Crenarchaea and AU bias in Euryarchaea. The Methanosarcinales have unique AU, the first base pair of D-stem, that differentiate these from tRNA$^{Tyr}$(GUA) of all other Archaea where GC is present. Base pair 2:71 is CG in general, but it is GC in Paero. G50:C64 of Nanoarchaea show identity with Crenarchaea and not with Euryarchaea where C50:G64 is present. The first base pair, CG in the A-arm is unique to tRNA$^{Tyr}$(GUA) and absent in other tRNA. This base pair is an identity element of tRNA$^{Tyr}$(GUA) for Tyr-RS

**Table 4.** List of introns in tRNA genes from Archaea.

| Archaea | Amino acids | | | | | | | | | | | | | | | | | | | | |
|---|---|---|---|---|---|---|---|---|---|---|---|---|---|---|---|---|---|---|---|---|---|
| | Ala | Arg | Asn | Asp | Cys | Gln | Glu | Gly | His | Ile | Leu | Lys | Phe | Pro | Ser | Thr | Trp | Tyr | Val | EMet1 | EMet2 | Imet |
| Paero | 20 | 16P56 | 13 | 21P3 | | 18P22 | 21P58 | 15 | 13 | | | | | 16 | | 20P29 | | 42 | 25 | 20P29 | 20P28 | 17 |
| | TGC | GCG; | GTT | GTC | | CTG | TTC | TCC | GTG | | | | | GGG; | | CGT; | | GTA; | TAC | CAT | CAT | CAT |
| | | 24P28 | | | | | | | | | | | | 15 | | 16P57 | | 20P29 | | | | |
| | | CCT; | | | | | | | | | | | | TGG | | GGT; | | GAC | | | | |
| | | 24P28 | | | | | | | | | | | | | | 20P29,20P59 | | | | | | |
| | | CCG | | | | | | | | | | | | | | TGT | | | | | | |
| Aper | | 44 | | 121 | 18 | | | | | | | 34 | | 44 | 36 | 49 | 37P29 | 39 | | 48 | 38 | |
| | | TCT | | GTC | GCA | | | | | | | CTT | | CGG; | CGA | CGT; | CCA | GTA | | CAT | CAT | |
| | | | | | | | | | | | | | | 37 | | 19P22 | | | | | | |
| | | | | | | | | | | | | | | GGG | | TGT | | | | | | |
| Ssolfa | | 15 | 14 | | 14, | | 17P21 | | | 12 | 16 | 22 | | 21 | 24 | 13 | 65 | 13 | | 16 | | 25 |
| | | TCT; | GTT | | 25P28 | | TTC; | | | GAT | CAA; | CTT; | | GGG | CGA | CGT; | CCA | GTA | | CAT | | CAT |
| | | 13 | | | GCA | | 16P20 | | | | 15 | 23 | | | | 15 | | | | | | |
| | | CCT | | | | | CTC | | | | TAA | TTT | | | | TGT | | | | | | |
| Stok | | 16 | | | 14 | | 17P20 | | | 16 | 13 | 27 | 17 | 18 | 26 | 24 | 48 | 19 | | 12 | 19 | 24 |
| | | TCG; | | | GCA | | CTC | | | GAT | CAA; | CTT; | GAA | GGG | CGA; | CGT; | CCA | GTA | | CAT | CAT | CAT |
| | | 13 | | | | | | | | | 16 | 25 | | | 11 | 16 | | | | | | |
| | | GCG; | | | | | | | | | CAG; | TTT | | | TGA | TGT | | | | | | |
| | | 24 | | | | | | | | | 16 | | | | | | | | | | | |
| | | CCT | | | | | | | | | TAG | | | | | | | | | | | |
| Afulg | | | | 35 | | | 16 | | | | 15 | | | | | | 62 | 24 | | | | |
| | | | | GTC | | | TTC | | | | CAA | | | | | | CCA | GTA | | | | |
| Halo | | | | | | | | | | | | | | | | | 102 | | | | 84 | |
| | | | | | | | | | | | | | | | | | CCA | | | | CAT | |
| Pf | | | | | | | | | | | | | | | | | 71 | | | 32 | | |
| | | | | | | | | | | | | | | | | | CCA | | | CAT | | |
| Phori | | | | | | | | | | | | | | | | | 71 | | | 31 | | |
| | | | | | | | | | | | | | | | | | CCA | | | CAT | | |
| Paby | | | | | | | | | | | | | | | | | 71 | | | 31 | | |
| | | | | | | | | | | | | | | | | | CCA | | | CAT | | |

| | | | | | | | | | | | | | |
|---|---|---|---|---|---|---|---|---|---|---|---|---|---|
| Mmari | | | | | | | | | | 89<br>CCA | | 40<br>CAT | |
| Mjan | | | | | | | | | | 33<br>CCA | | 35<br>CAT | |
| MmGoe | 20<br>TCG | | | | | | | | | 118<br>CCA | 42<br>GTA | 36<br>CAT | |
| Macet | 20<br>TCG | | | | | | | | | 118<br>CCA | 41<br>GTA | 36<br>CAT | |
| Mbar | | | | | | | | | | | | | |
| Mburt | | | | | | | | | | | | | |
| Tacid | | | | | 12<br>CCC | | | | | 69<br>CCA | 25<br>GTA | 14<br>CAT | |
| Tvol | | | | | 14<br>CCC | | | | | 69<br>CCA | 25<br>GTA | 14<br>CAT | |
| Pitor | | | | | | | | | | 69<br>CCA | 22<br>GTA | 15<br>CAT | |
| Facid | | | | | | | | | | | | | |
| Mther | | | | | | | | | 16;32P32<br>GGG | 68<br>CCA | | 35<br>CAT | |
| Mkan | | 27<br>GTT | 21<br>GCA | 21;31P20b<br>TTC | | | | 32<br>GAA | 15<br>GGG | 76<br>CCA | | 36<br>CAT | |
| NanoA | | | | 11P33<br>CUC | | 53<br>13P32;<br>25P71<br>ATG | 19<br>TAT | | | 13P30<br>CCA | 13<br>GTA | 66<br>CAT | 74P55;<br>29P68<br>CAT |

The introns present in the tRNA genes of Archaea are sorted out according to the amino acid encoded. Column indicates the genome name in abbreviated form (see Table 1 for full names). Each element of the table indicates the length of the intron and below the anticodon (as found in the tRNA gene, e.g. TAT, instead of UAU in the tRNA). The length of the non-canonical introns are written in the following format as ‘xxPyy’, where ‘xx’ is the intron length and ‘yy’ the position of the nucleotide preceding the intron. The canonical introns are simply denoted as numbers which indicate length of that intron. The tRNA genes harbouring two introns are separated by semicolon and below is the anticodon.

in addition to the anticodons. This observation is experimentally supported in Mjan.[65]

(xxi) *tRNA$^{Val}$*. MmGoe, Macet and Mbar have two identical copies of tRNA$^{Val}$(GAC). UA is present at 4:69 in tRNA$^{Val}$ of Halo and in tRNA$^{Val}$(GAC) of Mmari, but all other Archaea have CG. For archaeal tRNA$^{Val}$, the second and third nucleotides of anticodon, A35, C36 and the conserved, unique base pair of U11:A24 could be the identity elements for recognition by Val-RS. Additionally, the discriminator base A73 could be another identity element.

*Introns in archaeal tRNA genes*

A unique feature of archaeal tRNA genes is their ability to occasionally host introns at locations other than the usual position within the anticodon loop (between bases 37 and 38) of tRNA genes. These unusually located introns in archaeal tRNA genes were observed in 1987.[66] Identification of introns outside the anticodon loop was made two years later.[67] Recently, introns were discovered in AC-loop between bases 32 and 33 in pre-tRNA-Pro (GGG) of Mther and many other locations in tRNA genes of Paero. Introns at canonical positions are located by existing software (tRNAScan-SE and ARAGORN). Identification of non-canonical introns in archaeal tRNA genes continues to be a challenge. We record in Table 4 the non-canonical introns in archaeal tRNA genes identified by our in-house algorithm.

**Acknowledgements:** We acknowledge the useful communications from Christian Marck.

## References

1. Holley, R. W., Apgar, J., Everett, G. A., et al. 1965, Structure of ribonucleic acid, *Science*, **147**, 1462–1465.
2. Sugiura, M. and Wakasugi, T. 1989, Compilation and comparison of transfer RNA genes from tobacco chloroplasts, *CRC Critical Reviews in Plant Sciences*, **8**, 89–101.
3. Sugita, M., Shinozaki, K., and Sugiura, M. 1985, Tobacco chloroplast tRNA$^{Lys}$(UUU) gene contains a 2.5-kilobase-pair intron: an open reading frame and a conserved boundary sequence in the intron, *Proc. Natl Acad. Sci. USA*, **82**, 3557–3561.
4. Chattopadhyay, S., Sahoo, S., Kanner, W. A., and Chakrabarti, J. 2003, Pressures in Archaea l protein coding genes: a comparative study, *Comparative and Functional Genomics*, **4**, 56–65.
5. Fichant, G.A. and Burks, C. 1991, Identifying potential tRNA genes in genomic DNA sequences, *J. Mol. Biol.*, **220**, 659–671.
6. Pavesi, A., Conterio, F., Bolchi, A., Dieci, G., and Ottonello, S. 1994, Identification of new eukaryotic tRNA genes in genomic DNA databases by a multi-step weight matrix analysis of transcriptional control regions, *Nucleic Acids Res.*, **22**, 1247–1256.
7. el-Mabrouk, N. and Lisacek, F. 1996, Very fast identification of RNA motifs in genomic DNA: application to tRNA search in the yeast genome, *J. Mol. Biol.*, **264**, 46–55.
8. Lowe, T. M. and Eddy, S. R. 1997, tRNAscan-SE: a program for improved detection of transfer RNA genes in genomic sequence, *Nucleic Acids Res.*, **25**, 955–964.
9. Sagara, J., Shimizu, S., Kawabata, T., Nakamura, S., Ikeguchi, M., and Shimizu, K. 1998, The use of sequence comparison to detect 'identities' in tRNA genes, *Nucleic Acids Res.*, **26**, 1974–1979.
10. Gautheret, D. and Lambert, A. 2001, Direct RNA motif definition and identification from multiple sequence alignments using secondary structure profiles, *J. Mol. Biol.*, **313**, 1003–1011.
11. Laslett, D. and Canback, B. 2004, ARAGORN, a program to detect tRNA genes and tmRNA genes in nucleotide sequences, *Nucleic Acids Res.*, **32**, 11–16.
12. Steinberg, S., Misch, A., and Sprinzl, M. 1993, Compilation of tRNA sequences and sequences of tRNA genes, *Nucleic Acids Res.*, **21**, 3011–3015.
13. Sprinzl, M., Horn, C., Brown, M., Ioudovitch, A., and Steinberg, S. 1998, Compilation of tRNA sequences and sequences of tRNA genes, *Nucleic Acids Res.*, **26**, 148–153.
14. Marck, C. and Grosjean, H. 2002, tRNomics: analysis of tRNA genes from 50 genomes of Eukarya, Archaea and Bacteria reveals anticodon-sparing strategies and domain-specific features, *RNA*, **8**, 1189–1232.
15. Yusupov, M. M., Yusupova, G. Z., Baucom, A., et al. 2001, Crystal structure of the ribosome at 5.5 Å resolution, *Science*, **292**, 883–896.
16. Tamura, K., Nameki, N., Hasegawa, T., Shimizu, M., and Himeno, H. 1994, Role of the CCA terminal sequence of tRNA$^{Val}$ in aminoacylation with valyl-tRNA synthetase, *J. Biol. Chem.*, **269**, 22173–22177.
17. Hoagland, M. B., Zamecnik, P. C., and Stephenson, M. L. 1957, Intermediate reactions in protein synthesis, *Biochim. Biophys. Acta*, **24**, 215.
18. Himeno, H., Hasegawa, T., Ueda, T., Watnabe, K., and Shimizu, M. 1990, Conversion of aminoacylation specificity from tRNA$^{Tyr}$ to tRNA$^{Ser}$ *in vitro*, *Nucleic Acids Res.*, **18**, 6815–6819.
19. Ibba, M. and Söll, D. 2000, Aminoacyl-tRNA synthesis, *Annu. Rev. Biochem.*, **69**, 617–650.
20. Ibba, M. and Söll, D. 2001, The renaissance of aminoacyl-tRNA synthesis, *EMBO Reports*, **2**, 382–387.
21. Giege, R., Puglisi, J. D., and Florentz, C. 1993, tRNA structure and aminoacylation efficiency, *Nucleic Acid Res.*, **45**, 129–206.
22. Sherman, J. M., Rogers, M. J., and Soll, D. 1992, Competition of aminoacyl-tRNA synthetases for tRNA ensures the accuracy of aminoacylation, *Nucleic Acids Res.*, **20**, 2847–2852.
23. Yarus, M. 1972, Intrinsic precision of aminoacyl-tRNA synthesis enhanced through parallel systems of ligands, *Nat. New Biol.*, **239**, 106–108.
24. Rogers, M. J. and Söll, D. 1988, Discrimination between Glutaminyl-tRNA synthetase and Seryl-tRNA Synthetase involves nucleotides in the acceptor helix of tRNA, *Proc. Natl Acad. Sci. USA*, **85**, 6627–6631.

25. Hou, Y. M. and Schimmel, P. 1989, Modeling of *in vitro* kinetic parameters for the elaboration of transfer RNA identity *in vivo*, *Biochemistry*, **28**, 4942–4947.
26. Hou, Y. M., Motegi, H., Lipman, R. S., Hamann, C. S., and Shiba, K. 1999, Conservation of a tRNA core for aminoacylation, *Nucleic Acids Res.*, **27**, 4743–4750.
27. Steinberg, S. V., Leclerc, F., and Cedergren, R. 1997, Structural rules and conformational compensations in the tRNA L-form, *J. Mol. Biol.*, **266**, 269–282.
28. Zagryadskaya, E. I., Kotlova, N., and Steinberg, S. V. 2004, Key elements in maintainance of the tRNA L-shape, *J. Mol. Biol.*, **340**, 435–444.
29. Kim, S. H., Quickely, G. J., Suddath, F. L., McPherson, A., Sneden, D., and Kim, J. J. 1973, Three dimensional structure of yeast phenylalanine transfer RNA folding of the polypeptide chain, *Science*, **179**, 285–288.
30. Rorertus, R. D., Lander, J. E., Rhodes, J. T. F. D., Brown, R. S., Clark, B. F. C., and Klug, A. 1974, Structure of yeast phenylalanine tRNA at 3A° resolution, *Nature*, **250**, 546–551.
31. Rudinger, J., Blechschmidt, B., Ribeiro, S., and Sprinzl, M. 1994, Minimalist aminoacylated RNAs as efficient substrates for elongation factor Tu, *Biochemistry*, **33**, 5682–5688.
32. Fitz-Gibbon, S. T., Ladner, H., Kim, U. J., Stetter, K. O., Simon, M. I., and Miller, J. H. 2002, Genome sequence of the hyperthermophilic crenarchaeon *Pyrobaculum aerophilum*, *Proc. Natl Acad. Sci. USA*, **99**, 984–989.
33. Kawarabayasi, Y., Hino, Y., Horikawa, H., et al. 1999, Complete genome sequence of an aerobic hyperthermophilic crenarchaeon, *Aeropyrum pernix* K1, *DNA Res.*, **6**, 83–101, 145–152.
34. She, Q., Singh, R. K., Confalonieri, F., et al. 2001, The complete genome of the crenarchaeon *Sulfolobus solfataricus* P2, *Proc. Natl Acad. Sci. USA*, **98**, 7835–7840.
35. Kawarabayasi, Y., Hino, Y., Horikawa, H., et al. 2001, Complete genome sequence of an aerobic thermoacidophilic crenarchaeon, *Sulfolobus tokodaii* strain 7, *DNA Res.*, **8**, 123–140.
36. Klenk, H. P., Clayton, R. A., Tomb, J.-F., et al. 1997, The complete genome sequence of the hyperthermophilic, sulphate-reducing archaeon *Archaeoglobus fulgidus*, *Nature*, **390**, 364–370.
37. Ng, W. V., Kennedy, S. P., Mahairas, G. G., et al. 2000, Genome sequence of *Halobacterium* species NRC-1, *Proc. Natl Acad. Sci. USA*, **97**, 12176–12181.
38. Robb, F. T., Maeder, D. L., Brown, J. R., et al. 2001, Genomic sequence of hyperthermophile, *Pyrococcus furiosus*: implications for physiology and enzymology, *Meth. Enzymol.*, **330**, 134–157.
39. Kawarabayasi, Y., Sawada, M., Horikawa, H., et al. 1998, Complete sequence and gene organization of the genome of a hyper-thermophilic archaebacterium, *Pyrococcus horikoshii* OT3 (Supplement), *DNA Res.*, **5**, 147–155.
40. Cohen, G.N., Barbe, V., Flament, D., et al. 2003, An integrated analysis of the genome of the hyperthermophilic archaeon *Pyrococcus abyssi*, *Mol. Microbiol.*, **47**, 1495–1512.
41. Hendrickson, E. L., Kaul, R., Zhou, Y., et al. 2004, Complete genome sequence of the genetically tractable hydrogenotrophic methanogen *Methanococcus maripaludis*, *J. Bacteriol.*, **186**, 6956–6969.
42. Bult, C. J., White, O., Olsen, G. J., et al. 1996, Complete genome sequence of the methanogenic archaeon, *Methanococcus jannaschii*, *Science*, **273**, 1058–1073.
43. Deppenmeier, U., Johann, A., Hartsch, T., et al. 2002, The genome of *Methanosarcina mazei*: evidence for lateral gene transfer between Bacteria and Archaea, *J. Mol. Microbiol. Biotechnol.*, **4**, 453–461.
44. Galagan, J. E., Nusbaum, C., Roy, A., et al. 2002, The genome of *M. acetivorans* reveals extensive metabolic and physiological diversity, *Genome Res.*, **12**, 532–542.
45. Ruepp, A., Graml, W., Santos-Martinez, M. L., et al. 2000, The genome sequence of the thermoacidophilic scavenger *Thermoplasma acidophilum*, *Nature*, **407**, 508–513.
46. Kawashima, T., Amano, N., Koike, H., et al. 2000, Archaea l adaptation to higher temperatures revealed by genomic sequence of *Thermoplasma volcanium*, *Proc. Natl Acad. Sci. USA*, **97**, 14257–14262.
47. Futterer, O., Angelov, A., Liesegang, H., et al. 2004, Genome sequence of *Picrophilus torridus* and its implications for life around pH 0, *Proc. Natl. Acad. Sci. USA*, **101**, 9091–9096.
48. Smith, D. R., Doucette-Stamm, L. A., Deloughery, C., et al. 1997, Complete genome sequence of *Methanobacterium thermoautotrophicum* delta H: functional analysis and comparative genomics, *J. Bacteriol.*, **179**, 7135–7155.
49. Slesarev, A. I., Mezhevaya, K. V., Makarova, K. S., et al. 2002, The complete genome of hyperthermophile *Methanopyrus kandleri* AV19 and monophyly of Archaea l methanogens, *Proc. Natl. Acad. Sci. USA*, **99**, 4644–4649.
50. Waters, E., Hohn, M. J., Ahel, I., et al. 2003, The genome of *Nanoarchaeum equitans*: insights into early Archaea l evolution and derived parasitism, *Proc. Natl. Acad. Sci. USA*, **100**, 12984–12988.
51. Limmer, S., Reif, B., Ott, G., Arnold, L., and Sprinzl, M. 1996, NMR evidence for helix geometry modifications by a G-U wobble base pair in the acceptor arm of *E.coli* $tRNA^{Ala}$, *FEBS Lett*, **385**, 15–20.
52. Gabriel, K., Schneider, J., and McClain, W. H. 1996, Functional evidence for indirect recognition of G-U in $tRNA^{Ala}$ by alanyl-tRNA synthetases, *Science*, **271**, 195–197.
53. Kim, S. I. and Söll, D. 1998, Major identity element of glutamine tRNAs from *Bacillus subtilis* and *Escherichia coli* in the reaction with *B. subtilis* glutamyl-tRNA synthetase, *Mol. Cell.*, **8**, 459–465.
54. Sekine, S., Nureki, O., Tateno, M., and Yokoyama, S. 1999, The identity determinants required for the discrimination between $tRNA^{Glu}$ and $tRNA^{Asp}$ by glutamyl-tRNA synthetase from *Escherichia coli*, *Eur. J. Biochem.*, **261**, 354–360.
55. Madore, E., Florentz, C., Giegé, R., Sekine, S., Yokoyama, S., and Lapointe, J. 1999, Effect of modified nucleotides on *Escherichia coli* $tRNA^{Glu}$ structure and on its aminoacylation by glutamyl-tRNA synthetase, *Eur. J. Biochem.*, **266**, 1128–1135.
56. Tamura, K., Himeno, H., Ashara, H., Hasegawa, T., and Shmizu, M. 1992, *In vitro* study of *E.coli* $tRNA^{Arg}$ and

tRNA$^{Lys}$ identity elements, *Nucleic Acids Res.*, **20**, 2335–2339.
57. McClain, W. H., Foss, K., Jenkins, R. A., and Schneider, J. 1990, Nucleotides that determine *Escherichia coli* tRNA$^{Arg}$ and tRNA$^{Lys}$ acceptor identities revealed by analyses of mutant Opal and Amber suppressor tRNAs, *Proc. Natl. Acad. Sci. USA*, **87**, 9260–9264.
58. Mandal, N., Dev, M., Joseph, J., Dalluge, J., McCloskey, A., and Rajbhandary, U. L. 1996, Role of the three consecutive G:C base pairs conserved in the anticodon stem of initiator tRNAs in initiation of protein synthesis in *Escherichia coli*, *RNA*, **2**, 473–482.
59. Burke, B., Lipman, R. S. A., Shiba, K., Musier-Forsyth, K., and Hou, Y. M. 2001, Divergent adaptation of tRNA recognition by *Methanococcus jannaschii* Prolyl-tRNA synthetases, *J. Biol. Chem.*, **276**, 20286–20291.
60. Lipman, R. S. A., Sowers, K. R., and Hou, Y. M. 2000, Synthesis of Cysteinyl-tRNA$^{Cys}$ by a genome that lacks the normal cysteine-tRNA-Synthetase, *Biochemistry*, **39**, 7792–7798.
61. Lenhard, B., Orellana, O., Ibba, M., and Weygand-Durasevic, I. 1999, tRNA recognition and evolution of determinants in seryl-tRNA synthesis, *Nucleic Acids Res.*, **27**, 721–729.
62. Nagaoka, Y., Yokozawa, J., Umehara, T., Iwaki, J., et al. 2002, Molecular recognition of threonine tRNA by threonyl-tRNA synthetase from an extreme thermophilic archaeon, *Aeropyrum pernix* K1, *Nucleic Acids Res.*, **2**, 81–82.
63. Giege, R., Sissler, M., and Florentz, C. 1998, Universal rules and idiosyncratic features in tRNA identity, *Nucleic Acids Res.*, **26**, 5017–5035.
64. Rogers, M. J., Adachi, T., Inokuchi, H., and Söll, D. 1992, Switching tRNA$^{Gln}$ identity from glutamine to tryptophan, *Proc. Natl. Acad. Sci. USA*, **89**, 3463–3467.
65. Kobayashi, T., Nureki, O., Ishitani, R., et al. 2003, Structural basis for orthogonal tRNA specificities of tyrosyl-tRNA synthetases for genetic code expansion, *Nat. Struct. Biol.*, **10**, 425–432.
66. Wich, G., Leinfelder, W., and Böck, A. 1987, Genes for stable RNA in the extreme thermophile *Thermoproteus tenax*: introns and transcription signal, *EMBO J*, **6**, 523–528.
67. Kjems, J., Leffers, H., Olesen, T., and Garrett, R. A. 1989b, A unique tRNA intron in the variable loop of the extreme thermophile *Thermofilum pendens* and its possible evolutionary implications, *J. Biol. Chem.*, **264**, 17834–17837.